\documentstyle[11pt,paspconf]{article}

\begin{document}

\title{Cepheid Variables in the LMC and SMC}

\author{D.L.~Welch\altaffilmark{1},
        C.~Alcock\altaffilmark{2,3}, 
      R.A.~Allsman\altaffilmark{4},
        D.~Alves\altaffilmark{2,5,6},
      T.S.~Axelrod\altaffilmark{7},
      A.C.~Becker\altaffilmark{8},
      D.P.~Bennett\altaffilmark{2,3,9},
      K.H.~Cook\altaffilmark{2,3},
      K.C.~Freeman\altaffilmark{7},
        K.~Griest\altaffilmark{3,10},
      M.J.~Lehner\altaffilmark{3,10,11},
      D.W.~Kurtz\altaffilmark{12},
      S.L.~Marshall\altaffilmark{2,3},
        D.~Minniti\altaffilmark{2,3},    
      B.A.~Peterson\altaffilmark{7},
      M.R.~Pratt\altaffilmark{3,8,13},
      P.J.~Quinn\altaffilmark{14},
      A.W.~Rodgers\altaffilmark{5,15},
        A.~Rorabeck\altaffilmark{1},
        W.~Sutherland\altaffilmark{16},
        A.~Tomaney\altaffilmark{8}, and
        T.~Vandehei\altaffilmark{3,10}
      {\bf (The MACHO Collaboration)}
}

\altaffiltext{1}{Dept. of Physics \& Astronomy, McMaster University,
        Hamilton, Ontario, L8S 4M1 Canada
        E-mail: {\tt welch@physics.mcmaster.ca, rorabeck@glen-net.ca}}

\altaffiltext{2}{Lawrence Livermore National Laboratory, Livermore, CA 94550 \\
        E-mail: {\tt alcock, alves, dminniti, kcook, stuart@igpp.llnl.gov}}
 
\altaffiltext{3}{Center for Particle Astrophysics,
        University of California, Berkeley, CA 94720}
 
\altaffiltext{4}{Supercomputing Facility, Australian National University,
        Canberra, ACT 0200, Australia \\
        E-mail: {\tt robyn@macho.anu.edu.au}}

\altaffiltext{5}{Department of Physics, University of California,
        Davis, CA 95616 }

\altaffiltext{6}{Space Telescope Science Institute, 3700 San Martin Drive,
        Baltimore, MD 21218 \\
        E-mail: {\tt alves@stsci.edu}}
 
\altaffiltext{7}{Mt.~Stromlo and Siding Spring Observatories, Australian
        National University, Weston Creek, ACT 2611, Australia
        \ \ \ E-mail: {\tt tsa, kcf, peterson@mso.anu.edu.au}}
 
\altaffiltext{8}{Departments of Astronomy and Physics,
        University of Washington, Seattle, WA 98195 \\
        E-mail: {\tt becker, austin@astro.washington.edu}}

\altaffiltext{9}{Physics Department, University of Notre Dame, Notre 
        Dame, IN 46556 \\
        E-mail: {\tt bennett.27@nd.edu}}

\altaffiltext{10}{Department of Physics, University of California,
        San Diego, La Jolla, CA 92093 \\
        E-mail: {\tt kgriest, tvandehei@ucsd.edu }}

\altaffiltext{11}{Department of Physics, University of Sheffield,
        Sheffield, S3 7RH, U.K. \\
        E-mail: {\tt M.Lehner@sheffield.ac.uk}}

\altaffiltext{12}{Department of Astronomy, University of Cape Town, Rondebosch
        7701, South Africa \\
        E-mail: {\tt dkurtz@physci.uct.ac.za}}
 
\altaffiltext{13}{LIGO Project, MIT, Room 20B-145, Cambridge, MA 02139 \\
        E-mail: {\tt mrp@ligo.mit.edu}}

\altaffiltext{14}{European Southern Observatory, Karl-Schwarzchild Str. 2,
        D-85748, Garching, Germany \\
        E-mail: {\tt pjq@eso.org}}

\altaffiltext{15}{Deceased.}

\altaffiltext{16}{Department of Physics, University of Oxford, Oxford OX1
        3RH, U.K. \\
        E-mail: {\tt w.sutherland@physics.ox.ac.uk}}

\begin{abstract}
In this paper, we will review major new results regarding classical Cepheids,
in the Large Magellanic Cloud (LMC) and Small Magellanic Cloud (SMC). 
Specifically, we discuss recent work regarding multimode Cepheids and describe 
new observations of a W Vir star (HV 5756) and a Cepheid which are each in 
eclipsing binary systems. An additional interesting pulsating supergiant in
an eclipsing system is also identified. Ephemerides for eclipses for the
three systems are provided.
\end{abstract}


\section{MACHO Project Status}
The MACHO Project is a wide-field photometric survey whose primary goal
is to characterize the mass of non-luminous or under-luminous matter along
different lines-of-sight in the Milky Way galaxy. As of July 1998, observations
span 2100 nights with over 75,000 images archived. Data collection has taken
place essentially continuously since 1992, making this one of the most
extensive data sets for variable stars obtained in the history of variable
star astronomy. It is expected that similar observations will be obtained up
until at least 1999 December 31. A relatively recent review of the results
of the MACHO Project can be found in Cook {\it et al.} (1997).

\section{Classical Cepheids}
At present, the MACHO Project has identified 1767 classical and
first-overtone Cepheids in the LMC. EROS has reported a total of
290 Cepheids in the LMC. Both groups agree that the fraction of
first-overtone Cepheids is 30\%. At shorter periods, the fundamental
and first-overtone mode sequences are less distinct, suggesting the
possibility of a range in masses and/or ages for these stars. Such
a conclusion has been suggested by Bersier {\it et al.} (1998) and
Alcock {\it et al.} (1999a).

\section{Beat Cepheids}
Alcock {\it et al.} (1999b) report some of the results found by
Rorabeck (1997). The second-overtone lightcurve derived from 
1st/2nd-overtone beat Cepheid lightcurves is found to be very
sinusoidal. Beaulieu {\it et al.} (1997), Welch (1998), and
Rorabeck (1997) have reported beat Cepheids in the SMC.

\section{Eclipsing Cepheid Variables}

We have discovered three Cepheid variables in eclipsing binary
systems to date. Times of primary and secondary minima for
these systems are given in Tables 1, 2, and 3, respectively.

\begin{enumerate}
\item{{\sl HV 5756}} Our first reported discovery, HV 5756 (MACHO ID
78.6338.24) was reported in Welch {\it et al.} (1996). A total of
six primary minima have been observed at this writing (Dec 1998).
Furthermore, images from the Harvard College Observatory plate
archives have now been obtained to search for historical minima.
\item{{\sl 6.6454.5}} A finder chart with suitable comparison stars
is available at the URL:
$${\hbox{\sl http://wwwmacho.mcmaster.ca/EclCep/}}$$
Diluted lightcurve subtracted photometry is shown in Figure 1.
\item{{\sl 5.4763.71}} This system appears to contain a
4.69-day Cepheid with a slowly-varying amplitude and has an orbital
period of 123.9 days. It is possibly a low-mass Cepheid (in between
the tradition classifications of BL Her and W Vir variables). It
is located at equinox J2000.0 coordinates 05:09:59.2 -69:58:28.
Both primary and secondary minima are visible in the lightcurve
phased with the orbital period. The system is also remarkable in
that secondary minimum appears to take place at an orbital phase
of 0.55, indicating an eccentric orbit.
\end{enumerate}

These systems are likely just the tip of the iceberg, since the
hotter companions are usually more luminous than the pulsator and
hence are easily selected.

\vspace{-0.25truein}
\begin{center}
\begin{table}\scriptsize
\caption{Eclipse Predictions for HV 5756} \label{tbl-1}
\begin{tabular}{cc|cc}
\tableline
&&&\\[-0.1truein]
\multicolumn{2}{c|}{\bf Primary}& \multicolumn{2}{c}{\bf Secondary}\\ 
JD & UT & JD & UT \\
\tableline
&&&\\[-0.1truein]
{\tt 2451122.6}&{\tt 1998 Nov \phantom{0}5.1}&{\tt 2451332.6}&{\tt 1999 Jun \phantom{0}3.1}\\
{\tt 2451542.6}&{\tt 1999 Dec 30.1}&{\tt 2451752.6}&{\tt 2000 Jul 27.1}\\
{\tt 2451962.6}&{\tt 2001 Feb 22.1}&{\tt 2452172.6}&{\tt 2001 Sep 20.1}\\
{\tt 2452382.6}&{\tt 2002 Apr 18.1}&{\tt 2452592.6}&{\tt 2002 Nov 14.1}\\
{\tt 2452802.6}&{\tt 2003 Jun 12.1}&{\tt 2453012.6}&{\tt 2004 Jan \phantom{0}8.1}\\
\tableline
\end{tabular}
\end{table}
\end{center}

\vspace{-0.5truein}
\begin{center}
\begin{table}\scriptsize
\caption{Eclipse Predictions for 6.6454.5} \label{tbl-2}
\begin{tabular}{cc|cc}
\tableline
&&&\\[-0.1truein]
\multicolumn{2}{c|}{\bf Primary}& \multicolumn{2}{c}{\bf Secondary}\\
JD & UT & JD & UT \\
\tableline
&&&\\[-0.1truein]
{\tt 2450867.06}&{\tt 1998 Feb 22.56}&{\tt 2451065.81}&{\tt 1998 Sep \phantom{0}9.31}\\
{\tt 2451264.56}&{\tt 1999 Mar 27.06}&{\tt 2451463.31}&{\tt 1999 Oct 11.81}\\
{\tt 2451662.06}&{\tt 2000 Apr 27.56}&{\tt 2451860.81}&{\tt 2000 Nov 12.31}\\
{\tt 2452059.56}&{\tt 2001 May 30.06}&{\tt 2452258.31}&{\tt 2001 Dec 14.81}\\
{\tt 2452457.06}&{\tt 2002 Jul \phantom{0}1.56}&{\tt 2452655.81}&{\tt 2003 Jan 16.31}\\
{\tt 2452854.56}&{\tt 2003 Aug \phantom{0}3.06}&{\tt 2453053.31}&{\tt 2004 Feb 17.81}\\
\tableline
\end{tabular}
\end{table}
\end{center}

\vspace{-0.5truein}
\begin{center}
\begin{table}\scriptsize
\caption{Eclipse Predictions for 5.4763.71} \label{tbl-3}
\begin{tabular}{cc|cc}
\tableline
&&&\\[-0.1truein]
\multicolumn{2}{c|}{\bf Primary}& \multicolumn{2}{c}{\bf Secondary}\\
JD & UT & JD & UT \\
\tableline
&&&\\[-0.1truein]
{\tt 2451270.40}&{\tt 1999 Apr \phantom{0}1.90}&{\tt 2451338.54}&{\tt 1999 Jun \phantom{0}9.04}\\
{\tt 2451394.30}&{\tt 1999 Aug \phantom{0}3.80}&{\tt 2451462.44}&{\tt 1999 Oct 10.94}\\
{\tt 2451518.20}&{\tt 1999 Dec \phantom{0}5.70}&{\tt 2451586.34}&{\tt 2000 Feb 11.85}\\
{\tt 2451642.10}&{\tt 2000 Apr \phantom{0}7.60}&{\tt 2451710.24}&{\tt 2000 Jun 14.75}\\
{\tt 2451766.00}&{\tt 2000 Aug \phantom{0}9.50}&{\tt 2451834.14}&{\tt 2000 Oct 16.65}\\
{\tt 2451889.90}&{\tt 2000 Dec 11.40}&{\tt 2451958.04}&{\tt 2001 Feb 17.54}\\
{\tt 2452013.80}&{\tt 2001 Apr 14.30}&{\tt 2452081.94}&{\tt 2001 Jun 21.44}\\
{\tt 2452137.70}&{\tt 2001 Aug 16.20}&{\tt 2452205.84}&{\tt 2001 Oct 23.35}\\
{\tt 2452261.60}&{\tt 2001 Dec 18.10}&{\tt 2452329.74}&{\tt 2002 Feb 24.25}\\
{\tt 2452385.50}&{\tt 2002 Apr 21.00}&{\tt 2452453.64}&{\tt 2002 Jun 28.15}\\
\tableline
\end{tabular}
\end{table}
\end{center}

\begin{figure}
\vspace{2.25in}
\plotfiddle{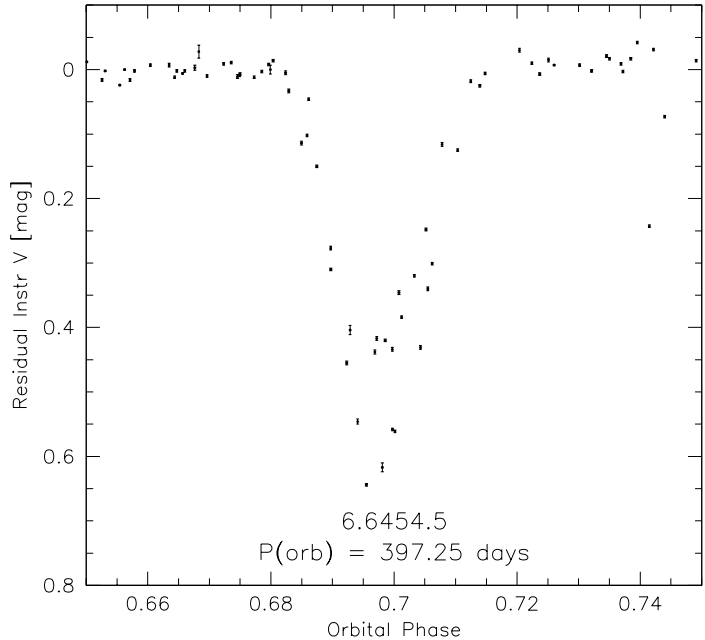}{0.0in}{0.0}{50}{50}{-150}{-100}
\caption{Five consecutive primary minima of the system containing
the 4.97-day classical Cepheid 6.6454.5 phased with an orbital 
period of 397.25 days. The scatter in the minima is not observational
in origin but is due to the removal of the contaminated Cepheid
lightcurve from the data. A full solution of the system is required
to remove the variation during eclipse properly. An epoch of mid-eclipse
is HJD 2449277.1. Primary eclipses typically last about 12 days.} \label{fig-1}
\end{figure}

\acknowledgments
Work performed at Lawrence Livermore National Laboratory (LLNL) is 
supported by the Department of Energy (DOE) under contract W7405-ENG-48. 
Work performed by the Center for Particle Astrophysics (CfPA) on the 
University of California campuses is supported in part by the Office 
of Science and Technology Centers of the National Science Foundation 
(NSF) under cooperative agreement AST-8809616. 
Work performed at MSO is supported by the Bilateral Science and Technology
Program of the Australian Department of Industry, Technology and Regional
Development.  DLW and AJR were supported, in part, by a Research Grant from
the Natural Sciences and Engineering Research Council of Canada (NSERC)
during this work.

\end{document}